%% file: main.tex
\documentclass[graybox]{svmult}
 
\usepackage{mathptmx}       
\usepackage{helvet}         
\usepackage{courier}        
\usepackage{type1cm}        
\usepackage{makeidx}         
\usepackage{graphicx}        
\usepackage{multicol}        
\usepackage[bottom]{footmisc}
\usepackage{amsmath}
\usepackage{amssymb}
\usepackage[mode=buildnew]{standalone}
\usepackage{booktabs}
\usepackage{multirow}
\usepackage{makecell}

\usepackage{xcolor}
\definecolor{mypink}{RGB}{187,0,86}
\definecolor{mediumturquoise}{RGB}{57, 193, 205}
\definecolor{myblue}{RGB}{0, 91, 127}

\usepackage{tikz}
\usetikzlibrary{matrix}
\usetikzlibrary{positioning}
\usetikzlibrary{arrows.meta}

\usepackage[final]{changes}
\definechangesauthor[name={Daria}, color=red]{DF}

\usepackage[justification=raggedright,singlelinecheck=false]{caption}
\usepackage{subcaption}
\makeindex             

\begin{document}

\title*{Machine learning methods for prediction of breakthrough curves in reactive porous media}
\titlerunning{ML for prediction of breakthrough curves}
\author{Daria Fokina, Pavel Toktaliev, Oleg Iliev and Ivan Oseledets}
\institute{Daria Fokina \at Fraunhofer ITWM, Kaiserslautern, Germany, Technical University Kaiserslautern, Kaiserslautern, Germany, \email{daria.fokina@itwm.fraunhofer.de}
\and Pavel Toktaliev \at Fraunhofer ITWM, Kaiserslautern, Germany, Technical University Kaiserslautern, Kaiserslautern, Germany, \email{pavel.toktaliev@itwm.fraunhofer.de}
\and Oleg Iliev \at Fraunhofer ITWM, Kaiserslautern, Germany, Bulgarian Academy of Sciences, Sofia, Bulgaria, \email{oleg.iliev@itwm.fraunhofer.de}
\and Ivan Oseledets \at Skolkovo Institute of Science and Technology, Moscow, Russia, Artificial Intelligence Research Institute, Moscow, Russia, \email{i.oseledets@skoltech.ru}
}
%
%
\maketitle

\abstract*{
Reactive flows in porous media play an important role in our life and are crucial for many industrial, environmental and biomedical applications. Very often the concentration of the species at the inlet is known, and the so-called breakthrough curves, measured at the outlet, are the quantities which could be measured or computed numerically. The measurements and the simulations could be time-consuming and expensive, and machine learning and Big Data approaches can help to predict breakthrough curves at lower costs. Machine learning (ML) methods, such as Gaussian processes and fully-connected neural networks, and a tensor method, cross approximation, are well suited for predicting breakthrough curves. In this paper, we demonstrate their performance in the case of pore scale reactive flow in catalytic filters.
}

\abstract{
Reactive flows in porous media play an important role in our life and are crucial for many industrial, environmental and biomedical applications. Very often the concentration of the species at the inlet is known, and the so-called breakthrough curves, measured at the outlet, are the quantities which could be measured or computed numerically. The measurements and the simulations could be time-consuming and expensive, and machine learning and Big Data approaches can help to predict breakthrough curves at lower costs. Machine learning (ML) methods, such as Gaussian processes and fully-connected neural networks, and a tensor method, cross approximation, are well suited for predicting breakthrough curves. In this paper, we demonstrate their performance in the case of pore scale reactive flow in catalytic filters.
}

\section{Introduction}

\replaced{Reactive flows in porous media}{Filtration processes} play an important role in our life and are crucial for \replaced{many industrial, environmental and biomedical applications}{the environment}. Therefore, it is important to understand how \replaced{the reactive transport}{these processes} work\added{s} and how one can  \replaced[id=DF]{control or}{\added{conrol or}} improve \replaced{it}{them}. \added{Note that chemical reactions occur at pore scale, however the measurements are usually done at macroscopic scale. Very often the concentration of the species at the inlet is known, and the so-called breakthrough curves are the quantities which could be measured or computed numerically. Breakthrough curves represent the concentration of the species at the flow outlet. In problems like parameter identification, optimizations, etc. \replaced[id=DF]{breakthrough}{breakthrogh} curves for many of the input parameters. The measurements and \replaced[id=DF]{simulations}{siulations} could be time-consuming and expensive, and machine learning and Big Data approaches can help to predict \replaced[id=DF]{breakthrough}{breaktghrough} curves at lower costs. Machine learning (ML) methods, such as Gaussian processes and fully-connected neural networks, and a tensor method, cross approximation, are well suited for predicting breakthrough curves. In this paper, we demonstrate their performance in the case of pore scale reactive flow in \replaced[id=DF]{catalytic}{catalitic} filters. The \replaced[id=DF]{computational}{computaional} geometry for a piece of the filtering membrane could come from 3D CT image, or can be  \replaced[id=DF]{generated by}{generatedby} proper software. The governing equation is the convection-diffusion-reaction equation, and linear and nonlinear reactions are considered. The computational experiments are followed by discussions about the \added[id=DF]{required} size of the training set, applicability and the performance of each of the methods for different flow regimes, namely diffusion  \replaced[id=DF]{dominated}{domiated} and reaction dominated. We expect that the considered methods will show similar performance for predicting breakthrough curves for other industrial or environmental problems, supposing \replaced[id=DF]{similar}{simiar} reactions and flow regimes are considered.} 

\deleted{
One example of a filtration system is a catalytic filter. Its efficiency can be described by a so-called breakthrough curve. The transport of species through a filter is modeled by a convection-diffusion-reaction equation. This model problem can be solved numerically. However, depending on the size of the geometry, this task may take several hours or days. When many such problems are to be solved, e.g. in case of parameter identification, this high-cost procedure can be replaced by machine learning methods, which predict values of the curve directly from input parameters of the filter. In this paper we present several of them in application to prediction of breakthrough curves w.r.t. filter parameters. The considered methods are machine learning (ML) methods, such as Gaussian processes and fully-connected neural networks, and a tensor method, cross approximation. The experiments are followed by the discussion about the applicability of each method for different use cases.
}

\section{\added[id=DF]{Problem formulation}}
\begin{figure}
	\centering
	\begin{subfigure}{0.45\textwidth}
		\includegraphics[width=\textwidth]{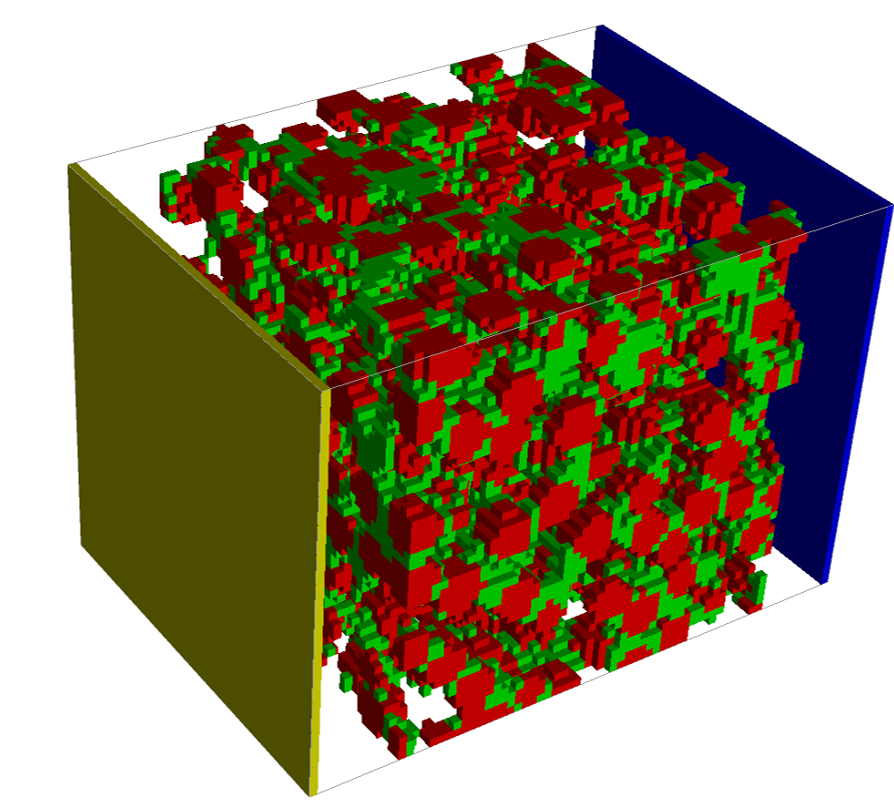}
		\caption{}
		\label{intro:geometry}
	\end{subfigure}
	\begin{subfigure}{0.45\textwidth}
		\includegraphics[width=\textwidth]{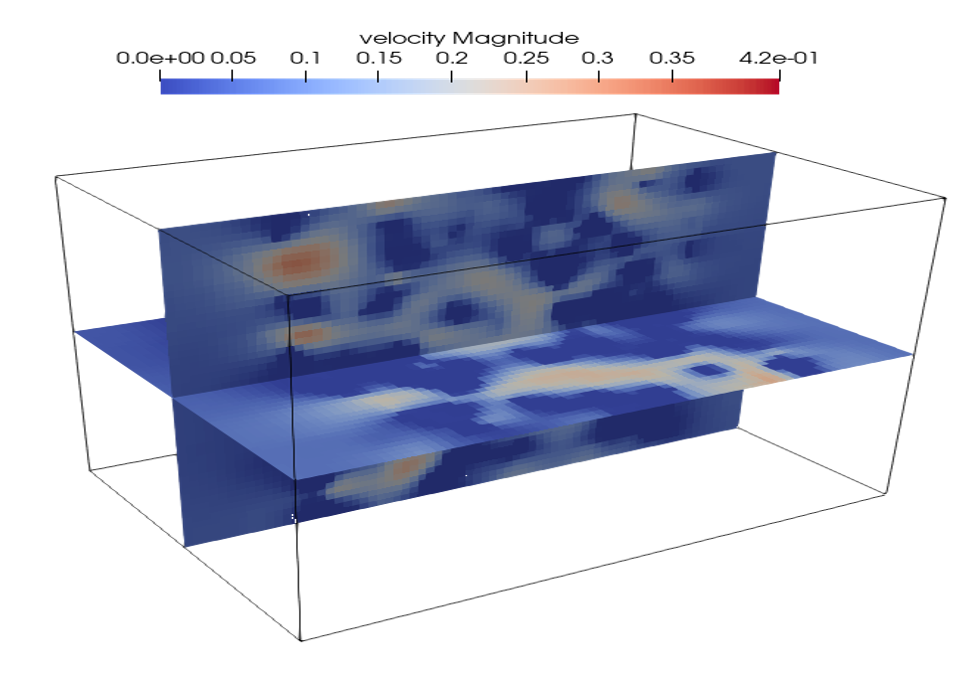}
		\caption{}
		\label{intro:vel}
	\end{subfigure}
	\caption{Considered geometry (a) and velocity field inside the media (b).}
	\label{intro:geom_vel}
\end{figure}

We consider at microscale a reactive flow through a piece of a catalytic filter membrane. A sample of the geometry can be \replaced{seen}{found} on Fig.~\ref{intro:geometry}. \added{The red color is reserved for the solid inert particles, the green color denotes the active washcoat particles, the microscale pores are transparent. There is no transport through the inert particles. Reactions occur within the washcoat \deleted[id=DF]{particle}, they occupy the domain indicated by $\Omega_w$. The resolved microscale pores occupy a domain denoted as $\Omega_f$. Thus, the computational domain is \replaced[id=DF]{$\Omega=\Omega_w \cup \Omega_f$}{$\Omega=\Omega_w \cup \Omega_w$}}. Mathematically, the process is modeled by the \added{the following} transport equation:
\begin{equation}
	c_t - D\;\Delta c + (\vec{v},\nabla c) + r(c) = 0,\quad x\in \Omega, t>0,
\end{equation}
$D$ is the diffusion coefficient \added{which is different in the pores and in the washcoat \deleted[id=DF]{particles}}, $v$ is the stationary velocity field \added{which is computed in advance, see below}, \added[id=DF]{$r(c) = 0,\, x\in\Omega_f$. For the reaction in the washcoat, we consider linear and nonlinear reactions. Further in the text we refer only to the processes in the washcoat.}
If \added{the} reaction is linear: $r(c) = kc$, then \added{in a dimensionless form} the equation can be written \added{as follows:} \deleted{in a dimensionless form:}
\begin{equation}
	c_t - \mu_{\mathrm{Da}}\; \Delta c + (\vec{v},\nabla c) + \mu_{\mathrm{Pe}}\, c = 0,
\end{equation}
where $\mu_{\mathrm{Da}} = \frac{kL}{D}$ is Damk\"ohler number, $\mu_{\mathrm{Pe}} = \frac{u_{in}L}{D}$ is Peclet number, $L$ is length of the \replaced{computational domain}{filter}, $u_{in}$ is the velocity at the inlet.

The initial and boundary conditions are \replaced{specified as follows}{the following}:
\begin{equation}
	\begin{gathered}
		c(x,0) = 0, \quad x\in\Omega,\\
		c(x,t) = 1, \quad x\in\Omega_{\mathrm{inlet}}, t>0,\\
		\frac{\partial c}{\partial \vec{n}} (x,t) = 0, \quad x\in\Omega_{\mathrm{outlet}}\cup \Omega_\mathrm{walls}.\\
	\end{gathered}
\end{equation}

Further we use notation: $\vec{\mu} = (\mu_\mathrm{Da},\mu_\mathrm{Pe})$.

The quantity of interest is the \added{time dependent} concentration \replaced{at}{on} the outlet, \replaced{i.e. the}{or} breakthrough curve:
\begin{equation}
	s(t;\mu) = s_\mu(t) = \int\limits_{\Omega_{\mathrm{outlet}}} c(x,t;\mu) dx.
\end{equation}

We use \emph{supervised} machine learning methods to approximate the dependency $s_\mu(t)$ as a function of the input parameters $\mu$. For that we need the values $s_\mu$ for a set of values of $\mu$. This can be done experimentally or compute\replaced{d}{r} numerically. In both cases we do not get \replaced[id=DF]{a function of time $s_\mu (t)$}{$s_\mu (t)$ \,---\, a function of time}, but vectors $\vec{s}_\mu$ for fixed moments of time. To obtain $\vec{s}_\mu$ we use PoreChem software \cite{porechem}, which provides us with the results of numerical simulations. Based on these numerical values we train a machine learning model and then choose the best configuration based on the results on new data, validation data, yet not seen by the model. The third set, the test set, is used to present the final results of the best models.

\section{Considered methods}
In the scope of this paper, we compare two machine learning methods: Gaussian processes and fully connected neural networks and one tensor method - cross approximation. In the following, we discuss these methods in detail. Please note \replaced[id=DF]{that for our problem other machine learning methods could be considered as well.}{there are a lot of other machine learning methods that could be considered, as well.}

\subsection{Machine learning methods}
The considered machine learning methods considered are the so-called \emph{regression} methods. This means that we have a function $f:\Omega\subset\mathbb{R}^n \rightarrow \mathbb{R}^m$, which we would like to approximate from several data points: $\mathcal{D}=(X, Y)$ which consists of pairs $\{(\vec{x}_i, \vec{y}_i=f(x_i))\}, i=1,\ldots,N$, $x_i$ is the input, and $y_i$ is the target value. The approximation is represented as a parameterized mapping $\hat{f}_\theta$, $\theta$ are the parameters. Normally optimal parameters $\theta^*$ that give the best fit to the data points are chosen by minimizing mean squared error (MSE):
\begin{equation}\label{mse_problem}
	l(\theta; \mathcal{D}) = \sum\limits_{(\vec{x}_i, \vec{y}_i) \in \mathcal{D}}\|\hat{f}_\theta(\vec{x}_i) - \vec{y}_i\|^2 \rightarrow \min\limits_\theta.
\end{equation}
This functional is often used in neural networks.

\subsubsection{Gaussian processes}
The first method to consider is a Gaussian process. For simplicity of the description, consider a function $f:\Omega \rightarrow \mathbb{R}$. Then set of given target values $Y = \vec{y}$ is a vector. This is a Bayesian method, which gives the conditional distribution $p(y_*|x_*, X, Y)$ of $y_* = f(x_*)$ at a new point $x_*$, given a training dataset $(X,Y)$. Consider a Gaussian process \replaced[id=DF]{$f(x) \sim GP(0, K)$}{$f(x) \sim \mathcal{GP}(0, K)$}, where $K$ is a kernel function, or simply a kernel. This means that for any vector $\vec{y}$, $y_i = f(x_i), \, i=1,\ldots,N$, has a normal distribution $\mathcal{N}(\vec{0}, K(X, X)), \, K(X,X)_{ij} = K(x_i, x_j), i,j = 1,\ldots,N$. When several values of $f$ are known for the set of points $X\subset \Omega$, the posterior distribution of $y_* = f(x_*)$ for the new point $x_* \in \Omega$ is the following:
\begin{equation}
	\begin{split}\label{gp_posterior}
		{y}_*| \vec{y}, X, x_* \sim \mathcal{N}(&K(x_*,X)K(X,X)^{-1} \vec{y},\\ &K(x_*,x_*) - K(x_*, X)K(X,X)^{-1}K(X,x_*)).
	\end{split}
\end{equation}
The approximation is the mean of this distribution, i.e. $\hat{f}(x_*) = K(x_*, X) K(X,X) \vec{y}$.

We consider three kernel functions:
\begin{itemize}
	\item Radial Basis Function (RBF) kernel: $K(x, z) = \exp\left(-\frac{d(x, z)^2}{2l^2}\right)$,
	\item Matern kernel: $K(x, z) = \frac{1}{\Gamma(\nu)2^{\nu-1}}\left(\frac{\sqrt{2\nu} d(x, z)}{l}\right)^lK_\nu\left(\frac{\sqrt{2\nu} d(x, z)}{l}\right)$,
	\item Rational quadratic (RQ) kernel: $K(x, z) = \left(\frac{1+d^2(x,z)}{2\alpha l^2}\right)^{-\alpha}$,
\end{itemize}
where $K_\nu$ is the modified Bessel function, $\Gamma(\nu)$ is the Gamma function, $d(x,z)$ is the Euclidean distance function.

\subsubsection{Neural networks}
Another \deleted{one of most used} \replaced[id=DF]{class of ML models}{ML method\deleted{s}} \added{which has \replaced[id=DF]{demonstrated a good}{demonstarted goof} efficiency} \replaced[id=DF]{are artificial neural networks (NNs)}{is \replaced{the}{a} neural networks (NN) \added{method}}. In our \replaced{considerations}{task} we use a classical fully-connected neural network. For this kind of neural network, the approximation is the following:
\begin{equation}
	f_\theta  = L_d \circ g \circ \ldots \circ L_2 \circ g \circ L_1,
\end{equation}
\added[id=DF]{where} $L_k(\vec{x}) = \vec{x}W_k + \vec{b}_k, \, W_k \in \mathbb{R}^{h_{k-1}\times h_k}, \vec{b}_k \in \mathbb{R}^{h_k}$, $\theta = \{(W_k, \vec{b}_k\}_{k=1}^d$, $h_0 = n,\, h_d = m$, $g$ is a nonlinear function, which is also called \added[id=DF]{an} activation function. The parameters are found by solving the minimization problem \ref{mse_problem}. This is usually done by stochastic gradient descent or its modifications. In this work we use Adam optimizer~\cite{adam}.
We experiment with different activation \replaced[id=DF]{functions}{function}:
\begin{itemize}
	\item Rectified linear unit (ReLU): $g(x) = \max(0,x)$,
	\item Leaky ReLU: $g(x) = \begin{cases} x, \; \text{if } x > 0\\ \alpha x,\; \text{if } x\le 0 \end{cases}, \; \alpha = 0.01$,
	\item Exponential linear unit (ELU):  $g(x) = \begin{cases} x, \; \text{if}\, x > 0\\ \alpha \exp(x-1),\; if x\le 0 \end{cases}, \; \alpha = 0.01$.
\end{itemize}

\subsection{Cross approximation}
\subsubsection{Discrete case}
Unlike the previous methods, cross approximation is a tensor method. As it follows from the name, it operates on tensors, \replaced[id=DF]{i.e.}{or} on $n$-dimensional matrices. We describe this tensor method, and then explain extension to continuous case. 

The input of the model is a vector $\mathbf{\mu}=(\mu_\mathrm{Da},\mu_\mathrm{Pe})$, each component of which is located \added{with}in \added{a closed} interval of interest: $\mu_\mathrm{Da}\in\gamma_1=[p_1^1, p_2^1],\, \mu_\mathrm{Pe}\in\gamma_1=[p_1^2, p_2^2]$. At the first step, the space of input parameters $\gamma_1\times\gamma_2$ is restricted to a $k\times k$ grid $A\times B$. This is done by uniformly discretizing each of intervals $\gamma_1$ and $\gamma_2$. The notation is the following:

\begin{tabular}{ccc}
	$\mu_\mathrm{Da}\in\{A_i\}$, & $A_i\in \gamma_1$, & $A_1 = p_1^1, A_k = p_2^1$\\
	$\mu_\mathrm{Pe}\in\{B_i\}$, & $B_i\in \gamma_2$, & $B_1 = p_1^2, A_k = p_2^2$ 
\end{tabular}

If we compute the curves $s_\mu$ for each element on the grid, we get a three-dimensional tensor $\tens{S}$\replaced[id=DF]{$\in \mathbb{R}^{k\times k\times N_T}$, where $N_T$ is the number of simulated time steps}{\added{, each element of which is a vector with dimension equal to the number of the time steps simuated}}. We suppose that $\tens{S}$ can be represented in a low-rank form using Tucker decomposition:
\begin{equation}
	\tens{S} = \tens{G} \times_1 U \times_2 V,    
\end{equation}
$G\in \mathbb{R}^{r_1\times r_2\times N_T},\, U\in\mathbb{R}^{r_1\times k},\, V\in\mathbb{R}^{r_2\times k}$. \added{The} \replaced{m}{M}ultiplication $\times_m$ is performed along the axis $m$. For  $\tens{C} = \tens{A}\times_m B$, where $\tens{A}\in \mathbb{R}^{d_1,\ldots, d_n}, B\in\mathbb{R}^{d_m, l}$, its elements \deleted{of} are expressed as:
\begin{equation}
	C_{i_1,\ldots, i_{m-1}, j, i_{m+1}, \ldots, i_n} = \sum\limits_{i_m} A_{i_1,\ldots,i_n}B_{i_m, j}.
\end{equation}

\begin{figure}
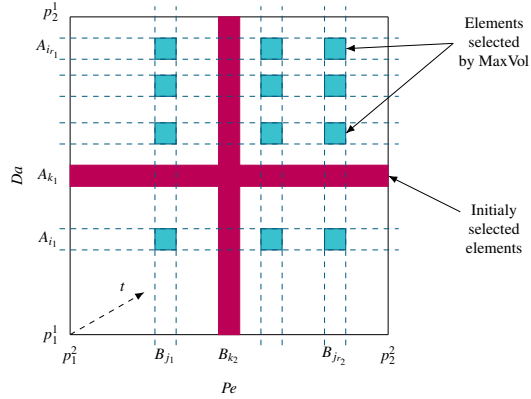

	\centering
	\includestandalone[width=.6\textwidth]{ca_tensor}
	\caption{Elements of tensor, required for cross approximation.}
	\label{methods:ca_tensor}
\end{figure}

Cross-approximation provides us with an algorithm to choose $U,\, V$ and $G$. 
Denote slices of $\tens{S}$ for some $k_1$ and $k_2$ as $P_1:=S[k_1,:.:]$, $P_2:=S[:,k_2,:]$. For this matrices we compute a singular value decomposition:
\begin{equation}
	P_1 = \hat{U}S_1\Phi_1; \, P_2 = \hat{V}S_2\Phi_2,
\end{equation}
with $U^T = \hat{U}[:,:r_1] \in \mathbb{R}^{k\times r_1}$ and $V^T = \hat{V}[:,:r_2]\in \mathbb{R}^{k\times r_2}$ are the leftmost singular vectors $r_1$ and $r_2$ respectively. To find $\tens{G}$ a MaxVol\cite{maxvol} algorithm is used. MaxVol is a greedy algorithm looking in a matrix for a submatrix of maximum volume (for a square matrix $\text{vol}(A) = |\text{det}(A)|$). This algorithm gives us a set of indices, for which we need \added{to} compute values of \replaced[id=DF]{$\tens{S}$}{$\tens{G}$}, to \replaced[id=DF]{get}{restore} \added{the} tensor $\tens{G}$, and, respectively, \added{to} reconstruct the whole tensor $\tens{S}$. \replaced[id=DF]{The}{All the} elements necessary for computing $S$ are \replaced{indicated in red and blue}{provided} on Fig.~\ref{methods:ca_tensor}

\subsubsection{Continuous case}
We consider the tensor $\tens{S}$ in the Tucker decomposition; this means that its elements can be represented as:
\begin{equation}
	S_{mnt} = \sum\limits_{i, j = 1}^{r_1, r_2} G_{ijt} U_{im} V_{jn},\quad m = 1,\ldots,k,\, n=1,\ldots,k,\, t=1,\ldots,N_T.
\end{equation}
The value of $u_{im}$ depends only on the value $A_i$ of \replaced[id=DF]{$\mu_{\mathrm{Da}}$}{$\mu_1$}, the value of $v_{jn}$ depends only on the value $B_j$ of \replaced[id=DF]{$\mu_{\mathrm{Pe}}$}{$\mu_2$}. In the continuous case, it would be written as:
\begin{equation}
	S(\mu_\mathrm{Da}, \mu_\mathrm{Pe}, t) = \sum\limits_{ij} G_{ij}(t) u_i(\mu_\mathrm{Da}) v_j(\mu_\mathrm{Pe}).
\end{equation}
We call $u_i(\mu_\mathrm{Da})$ and $v_j(\mu_\mathrm{Pe})$ as basis functions and find them by interpolating the vectors $U_i$ and $V_j$ correspondingly.

\section{\added{Computational} experiments} 
We conduct the experiments on a 3D geometry - a random packing of spheres. The size of \deleted[id=DF]{of} it is $50\times 50\times 50$ voxels. The length of the sample is $L=0.0001$ m, the diameter of the spheres is $l=6$ mkm, the velocity on the inlet \added[id=DF]{is} $u_{\mathrm{in}} = 0.047$, porosity is $\phi=0.741$. The geometry is shown on Fig.~\ref{intro:geometry} with the precomputed velocity field on Fig.~\ref{intro:vel}. To calculate the velocities, Geodict software \added{tool}  \cite{geodict} was used. \added[id=DF]{For each value of input parameters we compute concentration for 200 time steps ($N_T$) for $t\in[0.00001,0.002]$.}

We start with different regimes for the case of linear reaction and then consider the case of a strong nonlinear reaction. For the training of Gaussian processes and neural networks, we uniformly randomly sample the input parameters from the \replaced[id=DF]{domain}{interval} of interest. In each of the cases, the size of the training set for Gaussian processes is equal to 100 and for neural networks \added{it} is 600. For the cross approximation method we need parameter values for the middle cross-section of the parameter domain and also at positions found by MaxVol algorithm. For the evaluation of these three methods, we randomly sample 100 points for validation and separately 100 points for the test set. For all of these generated parameters, we numerically compute the breakthrough curves (target values for the models). In order to compare the results, we compare two errors:

\begin{align}
 e_{l_2} &= \sqrt{\frac{\sum\limits_{x\in\hat{\Omega}}\sum\limits_{i=1}^m (\hat{s}_i(x) - s_i(x))^2}{\sum\limits_{x\in\hat{\Omega}}\sum\limits_{i=1}^m s_i(x)^2}}\\
 e_{\max} &= \max\limits_{x\in\hat{\Omega}} \sqrt{\frac{\sum\limits_{i=1}^m (\hat{s}_i(x) - s_i(x))^2}{\sum\limits_{i=1}^m s_i(x)^2}}\\
\end{align}

\subsection{Linear reaction}
As it \replaced[id=DF]{was}{is} mentioned above, \replaced[id=DF]{we begin with a linear reaction $r(c)=k\,c$}{firstly a linear reaction $r(c) = k\;c$ is considered}. There are two regimes considered:
\begin{itemize}
	\item Diffusion dominated with $\mu_\textrm{Da} \in [0.01, 1.1],\, \mu_\textrm{Pe} \in [0.01, 1.1]$,
	\item Strong reaction with $\mu_\textrm{Da} \in [800., 1200.],\, \mu_\textrm{Pe} \in [0.05, 0.1]$.
\end{itemize}

\subsubsection{Diffusion dominated case}
\begin{figure}
	\centering
	\begin{subfigure}[t]{0.45\textwidth}
		\includegraphics[width=\textwidth]{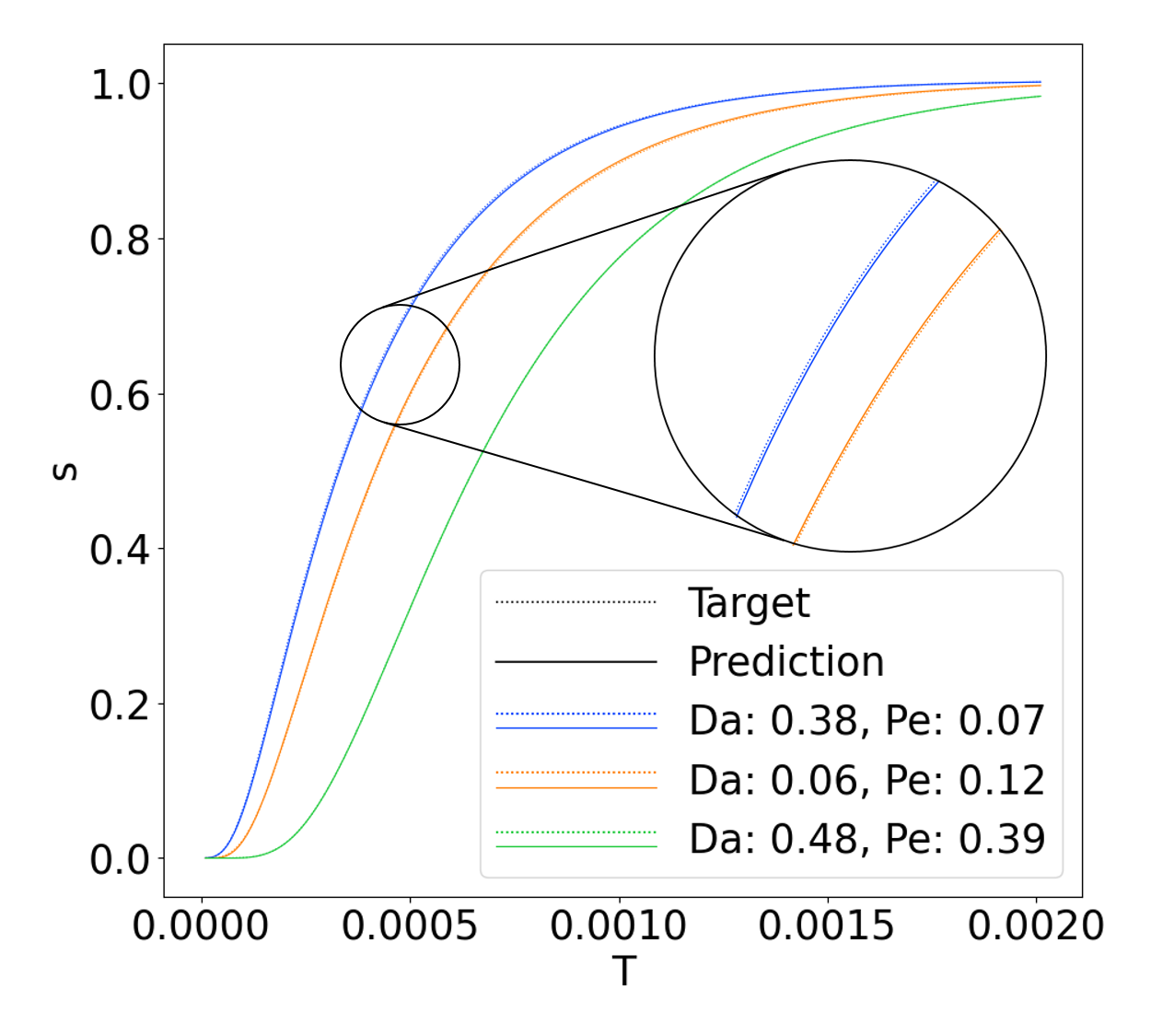}
		\label{lin:gp}
		\caption{}
	\end{subfigure}
	\begin{subfigure}[t]{0.45\textwidth}
		\includegraphics[width=\textwidth]{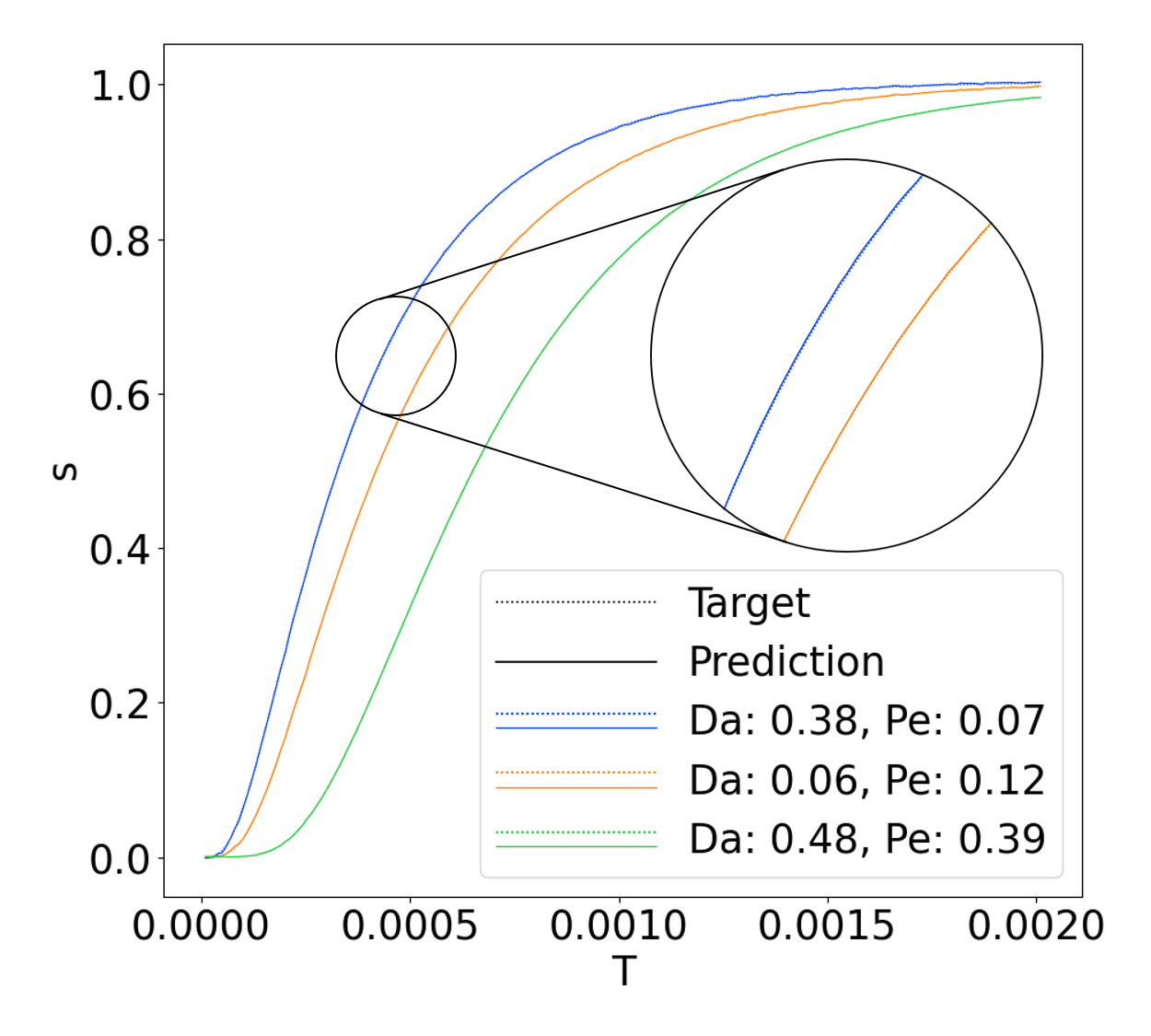}
		\caption{}
		\label{lin:nn}
	\end{subfigure}
	\begin{subfigure}[t]{0.45\textwidth}
		\includegraphics[width=\textwidth]{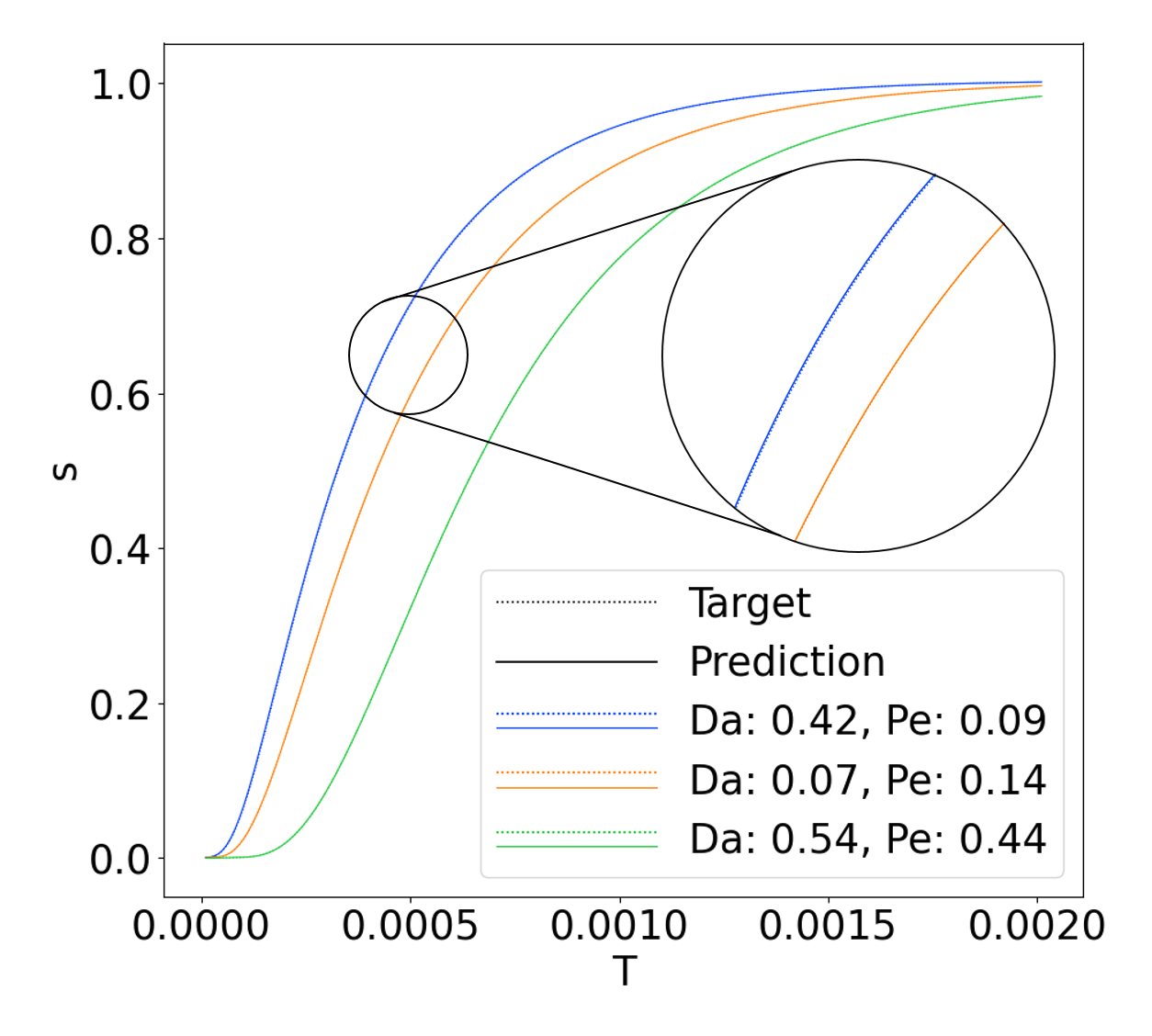}
		\caption{}
		\label{lin:ca}
	\end{subfigure}
	\caption{Linear reaction, diffusion dominated case. Breakthrough curve samples for target values and values predicted by: (a) Gaussian process, (b) neural network, (c) cross approximation.}
	\label{lin_diff:bcurves}
\end{figure}

The results for Gaussian processes for different kernels are \replaced[id=DF]{provided}{provides} in Table~\ref{lin_diff:errs_gp}, for neural networks for different activation functions are in Table~\ref{lin_diff:errs_nn}, for cross approximation \,---\, in Table~\ref{lin_diff:errs_ca}. Table~\ref{lin_diff:errs} provides the test errors for each method. Fig.~\ref{lin_diff:bcurves} compares breakthrough curves predicted by the models (solid line) with numerically computed target values (dotted line) for several values of input parameters $(\mu_\textrm{Da}, \mu_\textrm{Pe})$. The predicted values are provided for the best configuration of each method. The difference between predicted and target values is not seen by a human eye. On a zoomed-in figure a difference can be noticed, though it is also \replaced[id=DF]{very}{quite} small.

As one can see, the best result is achieved by a neural network with ELU activation function. The cross approximation gives a similar result, but with a smaller amount of training data.
\begin{table}[]
	\caption{Linear reaction, diffusion dominated case. Results for Gaussian processes for different kernels.}
	\label{lin_diff:errs_gp}
	
	\begin{tabular}{p{1.7cm}p{1.7cm}p{1.7cm}p{1.6cm}p{1.6cm}}
		\hline\noalign{\smallskip}
		\multirow{2}{*}{Kernel} & \multicolumn{2}{c}{Train error} & \multicolumn{2}{c}{Validation error}                \\
		\noalign{\smallskip}\cline{2-5}\noalign{\smallskip}
		& $e_{l_2}$             & $e_{\max}$           & $e_{l_2}$            & $e_{\max}$             \\
		\noalign{\smallskip}\svhline\noalign{\smallskip}
		RBF                     & $0.00161$            & $0.00489$            & $0.04281$            & $0.22586$          \\
		Matern                  & $1.0\times 10^{-7}$  & $5.96\times 10^{-7}$ & $\mathbf{0.04777}$   & $\mathbf{0.20501}$ \\
		RQ                      & $1.67\times 10^{-6}$ & $8.71\times 10^{-6}$ & $0.06546$            & $0.33635$          \\
		\noalign{\smallskip}\hline\noalign{\smallskip}
	\end{tabular}
\end{table}

\begin{table}[]
	\caption{Linear reaction, diffusion dominated case. Results for neural networks for activation functions.}
	\label{lin_diff:errs_nn}
	
	\begin{tabular}{p{2cm}p{1.6cm}p{1.6cm}p{1.6cm}p{1.6cm}}
		\hline\noalign{\smallskip}
		\multirow{2}{*}{\makecell{ Activation\\ function}} & \multicolumn{2}{c}{Train error} & \multicolumn{2}{c}{Validation error}   \\
		\noalign{\smallskip}\cline{2-5}\noalign{\smallskip}					  		 
							  & $e_{l_2}$      & $e_{\max}$    & $e_{l_2}$        & $e_{\max}$    \\
		\noalign{\smallskip}\svhline\noalign{\smallskip}
		ReLU                                   & $0.00011$     & $0.00040$        & $0.00166$             & $0.00905$          \\
		Leaky ReLU                             & $0.00038$     & $0.00136$        & $0.00248$             & $0.01771$          \\
		ELU                                    & $0.00084$     & $0.00475$        & $\mathbf{0.00113}$    & $\mathbf{0.00428}$ \\
	\noalign{\smallskip}\hline\noalign{\smallskip}
	\end{tabular}
\end{table}

\begin{table}[]
	\caption{Linear reaction, diffusion dominated case. Results for cross-approximation and different interpolation methods.}
	\label{lin_diff:errs_ca}
	
	\begin{tabular}{p{2.3cm}p{1.6cm}p{1.6cm}p{1.6cm}p{1.6cm}}
		\hline\noalign{\smallskip}
		\multirow{2}{*}{Interpolation} & \multicolumn{2}{c}{Train error} & \multicolumn{2}{c}{Validation error}   \\
		\noalign{\smallskip}\cline{2-5}\noalign{\smallskip}
							           & $e_{l_2}$     & $e_{\max}$    & $e_{l_2}$             & $e_{\max}$    \\
		\noalign{\smallskip}\svhline\noalign{\smallskip}
		Polynomial                     & $0.01026$     & $0.03127$     & $0.01041$             & $0.03135$          \\
		Piecewise linear               & $0.00210$     & $0.01820$     & $\mathbf{0.00380}$    & $\mathbf{0.01974}$ \\
		\noalign{\smallskip}\hline\noalign{\smallskip}
	\end{tabular}
\end{table}

\begin{table}[]
	\caption{Linear reaction, diffusion dominated case. Test errors}
	\label{lin_diff:errs}
	
	\begin{tabular}{p{3cm}p{1.6cm}p{1.6cm}}
		\hline\noalign{\smallskip}
		\multirow{2}{*}{Method} & \multicolumn{2}{c}{Test error}                   \\
		\noalign{\smallskip}\cline{2-3}\noalign{\smallskip}
								& $e_{l_2}$			   & $e_{\max}$        \\
		\noalign{\smallskip}\svhline\noalign{\smallskip}
		Gaussian process        & $0.01716$            & $0.13296$            \\
		Neural network          & $0.00055$            & $0.00126$ \\
		Cross approximation     & $0.00046$            & $0.00219$  \\
		\noalign{\smallskip}\hline\noalign{\smallskip}
	\end{tabular}
\end{table}

\subsubsection{Strong reaction}
In a case of \added[id=DF]{a} strong \added{linear} reaction (large Da number) the results are presented in \replaced[id=DF]{Table}{Tables}~\ref{lin_large_da:errs_gp} for Gaussian processes, Table~\ref{lin_large_da:errs_nn} for neural networks and Table~\ref{lin_large_da:errs_ca} for cross approximation. In Table~\ref{lin_large_da:errs} test errors are presented. We also plot the breakthrough curve samples, numerically computed and predicted by the considered methods.

\begin{figure}
	\centering
	\begin{subfigure}[t]{0.45\textwidth}
		\includegraphics[width=\textwidth]{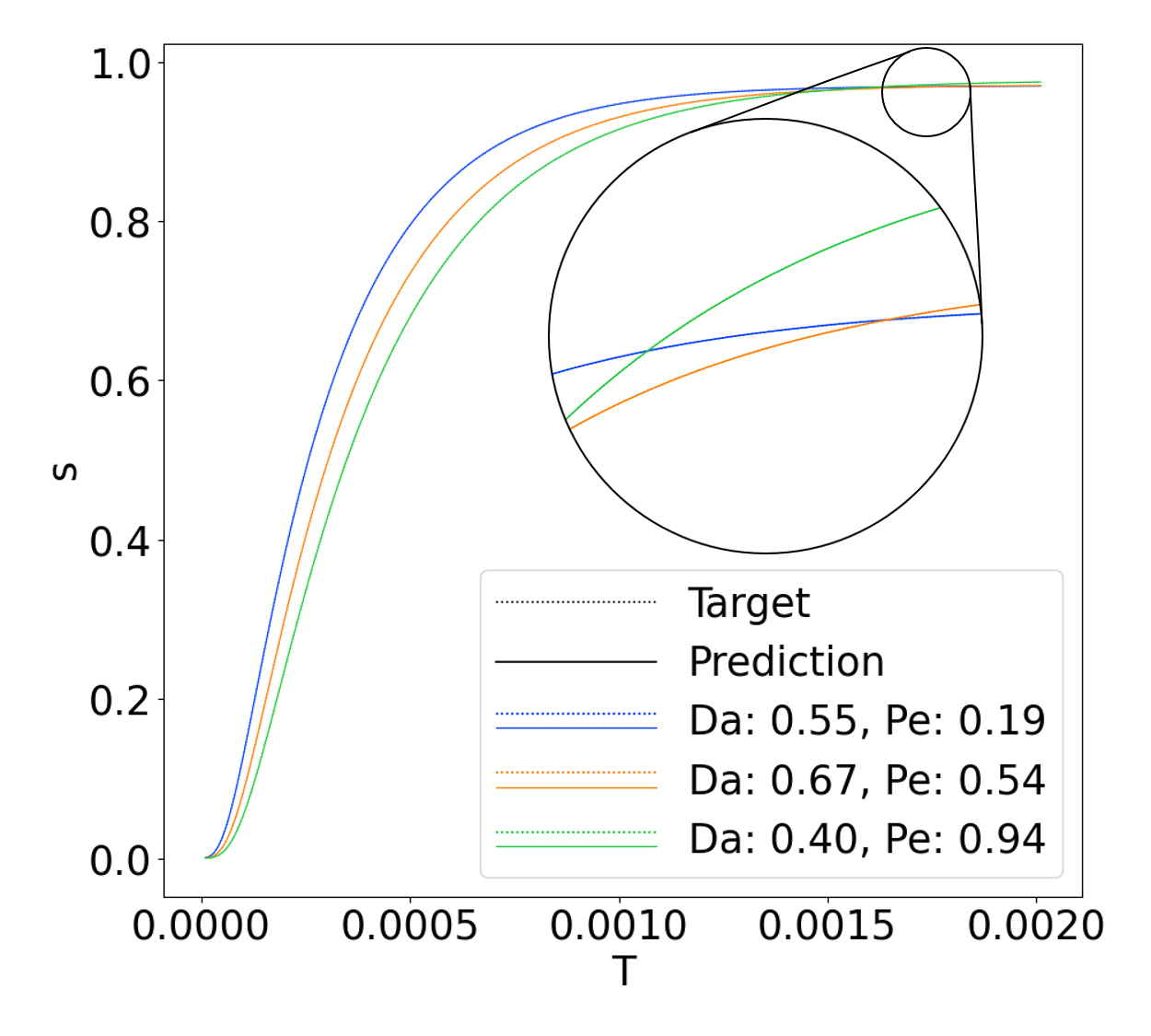}
		\label{lin_large_da:gp}
		\caption{}
	\end{subfigure}
	\begin{subfigure}[t]{0.45\textwidth}
		\includegraphics[width=\textwidth]{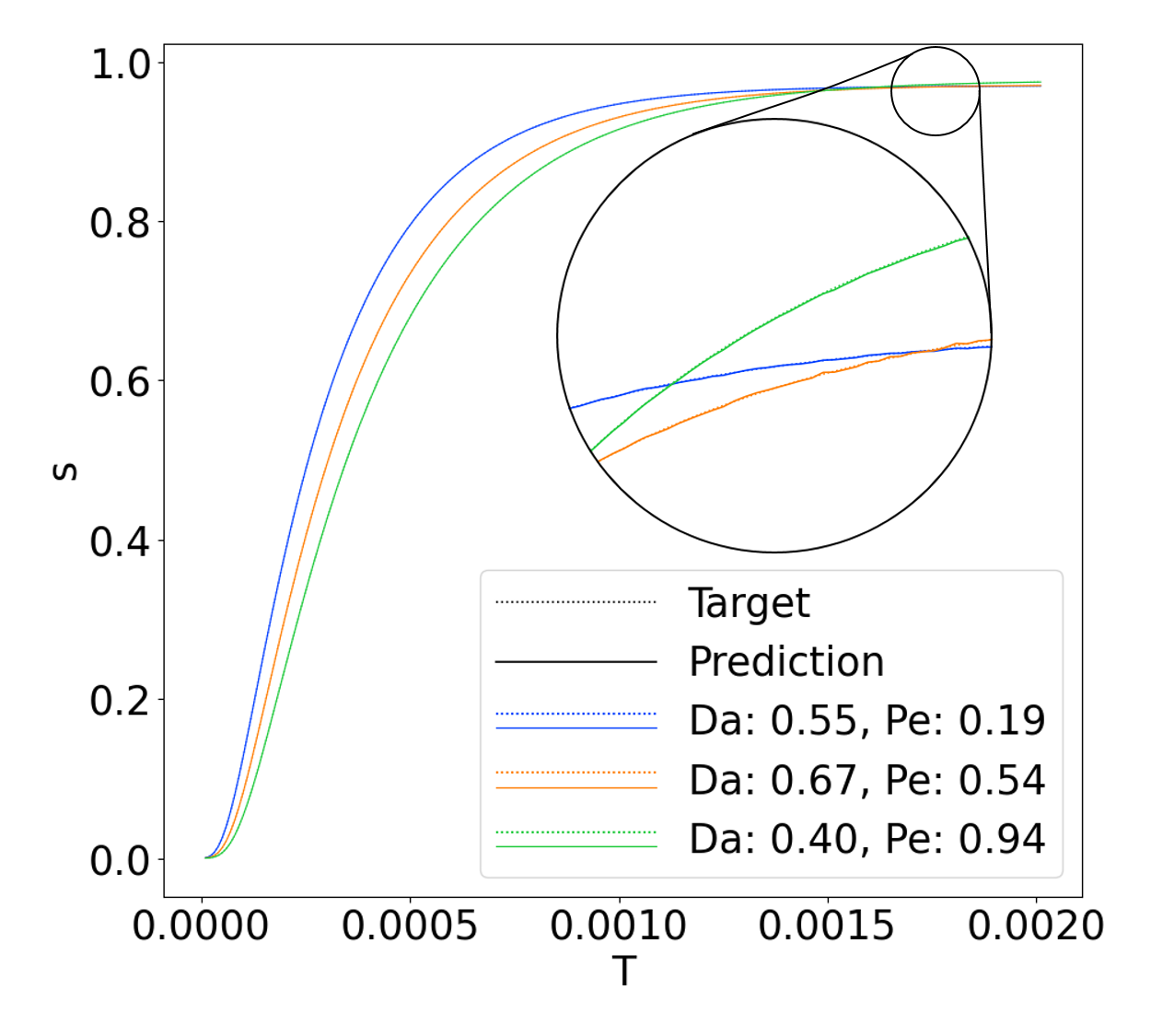}
		\caption{}
		\label{lin_large_da:nn}
	\end{subfigure}
	\begin{subfigure}[t]{0.45\textwidth}
		\includegraphics[width=\textwidth]{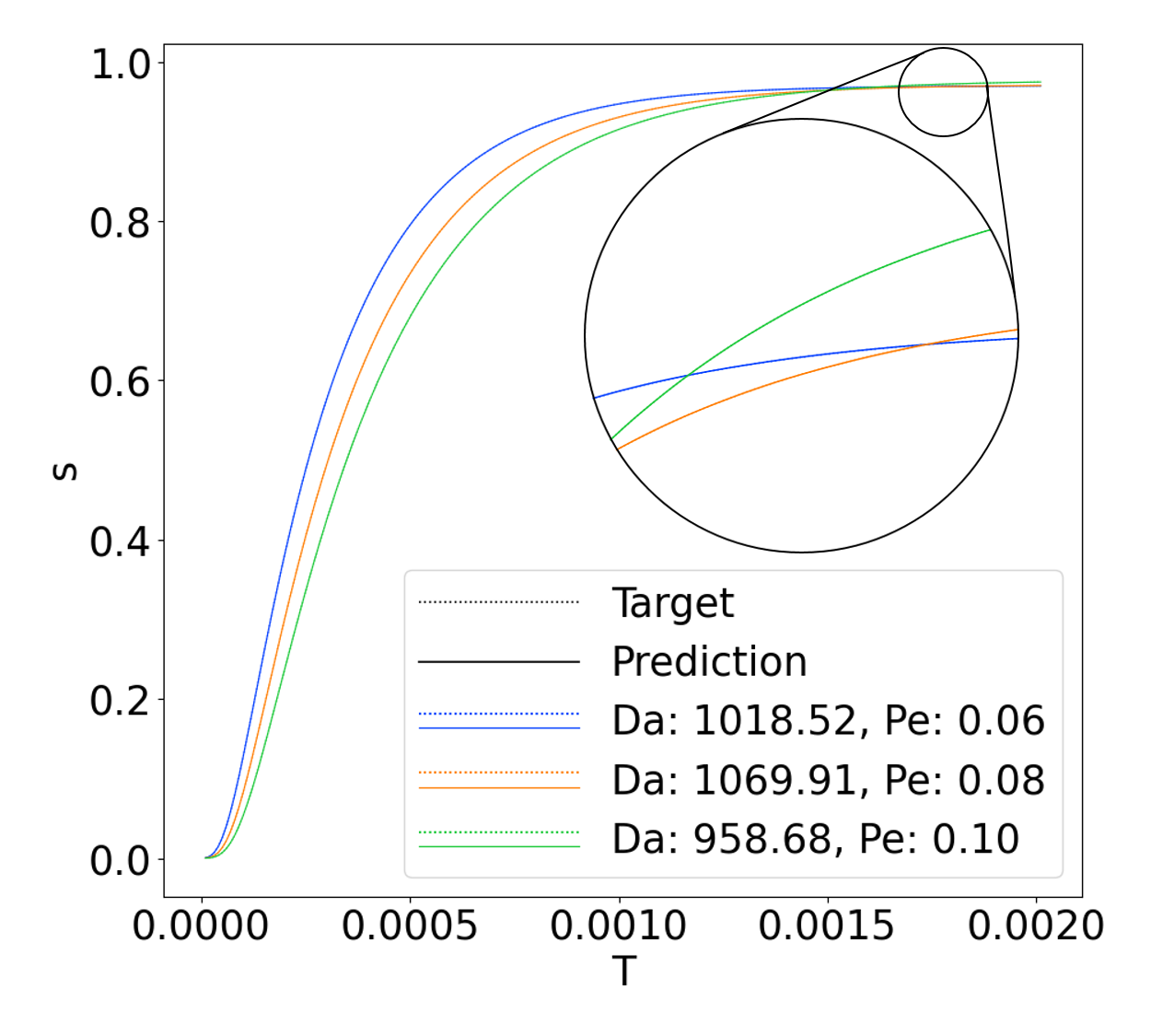}
		\caption{}
		\label{lin_large_da:ca}
	\end{subfigure}
	\caption{Linear strong reaction. Breakthrough curve samples for target values and values predicted by: (a) Gaussian process, (b) neural network, (c) cross approximation.}
	\label{lin_large_da:bcurves}
\end{figure}

In this case, the best performance shows cross-approximation + polynomial interpolation. For Gaussian processes with rational quadratic kernel the error is bigger, however the difference to cross approximation is minor. Neural networks have the worst result, although they use the most of the data. On the Fig.~\ref{lin_large_da:bcurves} we can notice the difference between the values predicted by the model and and numerically simulated data for neural networks, but this difference is noticed only at a closer look at the figure.
\begin{table}[]
	\caption{Linear strong reaction. Results for Gaussian processes for different kernels.}
	\label{lin_large_da:errs_gp}
	
	\begin{tabular}{p{1.7cm}p{1.7cm}p{1.7cm}p{1.7cm}p{1.7cm}}
		\hline\noalign{\smallskip}
		\multirow{2}{*}{Kernel} & \multicolumn{2}{c}{Train error}            & \multicolumn{2}{c}{Validation error}                          \\
		\noalign{\smallskip}\cline{2-5}\noalign{\smallskip}
								& $e_{l_2}$     	   & $e_{\max}$          & $e_{l_2}$ 					   & $e_{\max}$           \\
		\noalign{\smallskip}\svhline\noalign{\smallskip}
		RBF                     & $6.79\times 10^{-6}$ & $3.57\times 10^{-5}$ & $9.32\times 10^{-6}$           & $4.04\times 10^{-5}$          \\
		Matern                  & $1.83\times 10^{-5}$ & $0.00017$            & $3.97\times 10^{-5}$           & $0.00028$                     \\
		RQ                      & $2.05\times 10^{-6}$ & $9.41\times 10^{-6}$ & $\mathbf{8.37\times 10^{-6}}$  & $\mathbf{7.35\times 10^{-5}}$ \\
		\noalign{\smallskip}\hline\noalign{\smallskip}
	\end{tabular}
\end{table}

\begin{table}[]
	\caption{Linear strong reaction. Results for neural networks for activation functions.}
	\label{lin_large_da:errs_nn}
	
	\begin{tabular}{p{2cm}p{1.7cm}p{1.6cm}p{1.6cm}p{1.6cm}}
		\hline\noalign{\smallskip}
		\multirow{2}{*}{\makecell{ Activation\\ function}} & \multicolumn{2}{c}{Train error}   & \multicolumn{2}{c}{Validation error}    \\
		\noalign{\smallskip}\cline{2-5}\noalign{\smallskip}
											   & $e_{l_2}$     	      & $e_{\max}$  & $e_{l_2}$     	 & $e_{\max}$      \\
		\noalign{\smallskip}\svhline\noalign{\smallskip}
		ReLU                                   & $0.00297$            & $0.00661$   & $0.00322$          & $0.00573$           \\
		Leaky ReLU                             & $6.14\times 10^{-5}$ & $0.00016$   & $\mathbf{0.00055}$ & $\mathbf{0.00011}$  \\
		ELU                                    & $0.00109$            & $0.00234$   & $0.001163$         & $0.00186$           \\
		\noalign{\smallskip}\hline\noalign{\smallskip}
	\end{tabular}
\end{table}

\begin{table}[]
	\caption{Linear strong reaction. Results for cross-approximation and different interpolation methods.}
	\label{lin_large_da:errs_ca}
	
	\begin{tabular}{p{2.3cm}p{1.7cm}p{1.7cm}p{1.7cm}p{1.7cm}}
		\hline\noalign{\smallskip}
		\multirow{2}{*}{Interpolation} & \multicolumn{2}{c}{Train error}            & \multicolumn{2}{c}{Validation error}                  \\
		\noalign{\smallskip}\cline{2-5}\noalign{\smallskip}
									   & $e_{l_2}$     	      & $e_{\max}$           & $e_{l_2}$     	   			   & $e_{\max}$ \\
		\noalign{\smallskip}\svhline\noalign{\smallskip}
		Polynomial                     & $2.24\times 10^{-6}$ & $4.24\times 10^{-6}$ & $\mathbf{2.20\times 10^{-6}}$   & $\mathbf{3.48\times 10^{-6}}$          \\
		Piecewise linear               & $4.48\times 10^{-6}$ & $9.92\times 10^{-6}$ & $4.32\times 10^{-6}$            & $8.86\times 10^{-6}$ \\
		\noalign{\smallskip}\hline\noalign{\smallskip}
	\end{tabular}
\end{table}

\begin{table}[]
	\caption{Linear strong reaction. Test errors}
	\label{lin_large_da:errs}
	
	\begin{tabular}{p{3cm}p{1.7cm}p{1.7cm}}
		\hline\noalign{\smallskip}
		\multirow{2}{*}{Kernel} & \multicolumn{2}{c}{Test error}                   \\
		\noalign{\smallskip}\cline{2-3}\noalign{\smallskip}
								& $e_{l_2}$     	     & $e_{\max}$              \\
		\noalign{\smallskip}\svhline\noalign{\smallskip}
		Gaussian process        & $8.3\times 10^{-6}$    & $4.13\times 10^{-5}$            \\
		Neural network          & $8.83\times 10^{-5}$   & $0.00027$ \\
		Cross approximation     & $2.11\times 10^{-6}$   & $4.03\times 10^{-6}$  \\
		\noalign{\smallskip}\hline\noalign{\smallskip}
	\end{tabular}
\end{table}

\subsection{Nonlinear reaction} 
Here we consider \added{a} nonlinear reaction 
\begin{equation}
	r(c) = -\frac{k\cdot c}{(1+k_{\text{inhib}} c)^2},
\end{equation}
$k_{\text{inhib}} = 4 m^3/mol$.

Like in the case of the \added{the} linear reaction, we introduce \added{here in }the same way \deleted{here} $Da$ and $Pe$ numbers. The parameter intervals for this case are the same as for \added{the} strong linear reaction: $Da \in [800., 1200.],\, Pe\in [0.05, 0.1]$. 

\begin{figure}
	\centering
	\begin{subfigure}[t]{0.45\textwidth}
		\includegraphics[width=\textwidth]{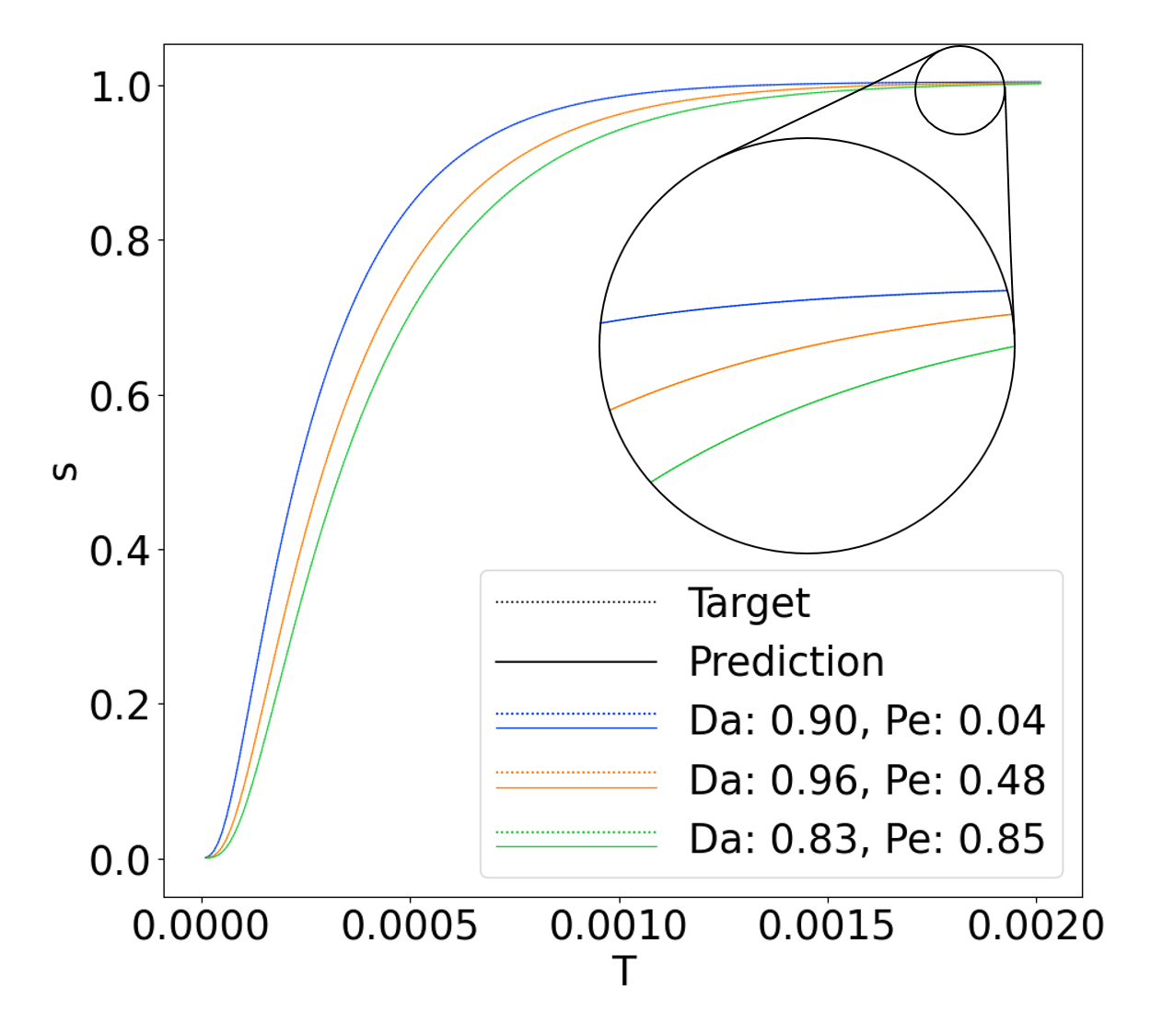}
		\label{nonlin:gp}
		\caption{}
	\end{subfigure}
	\begin{subfigure}[t]{0.45\textwidth}
		\includegraphics[width=\textwidth]{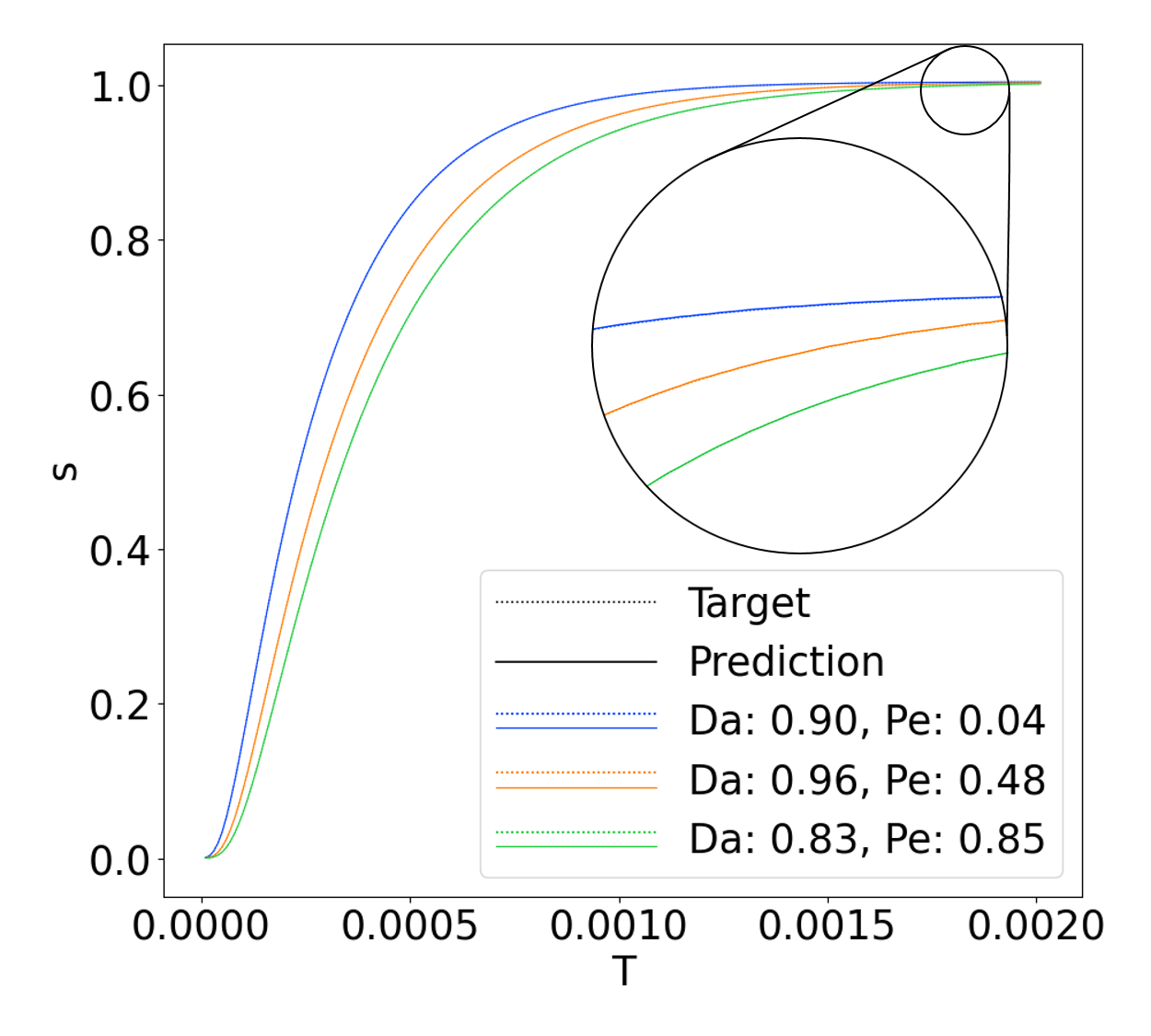}
		\caption{}
		\label{nonlin:nn}
	\end{subfigure}
	\begin{subfigure}[t]{0.45\textwidth}
		\includegraphics[width=\textwidth]{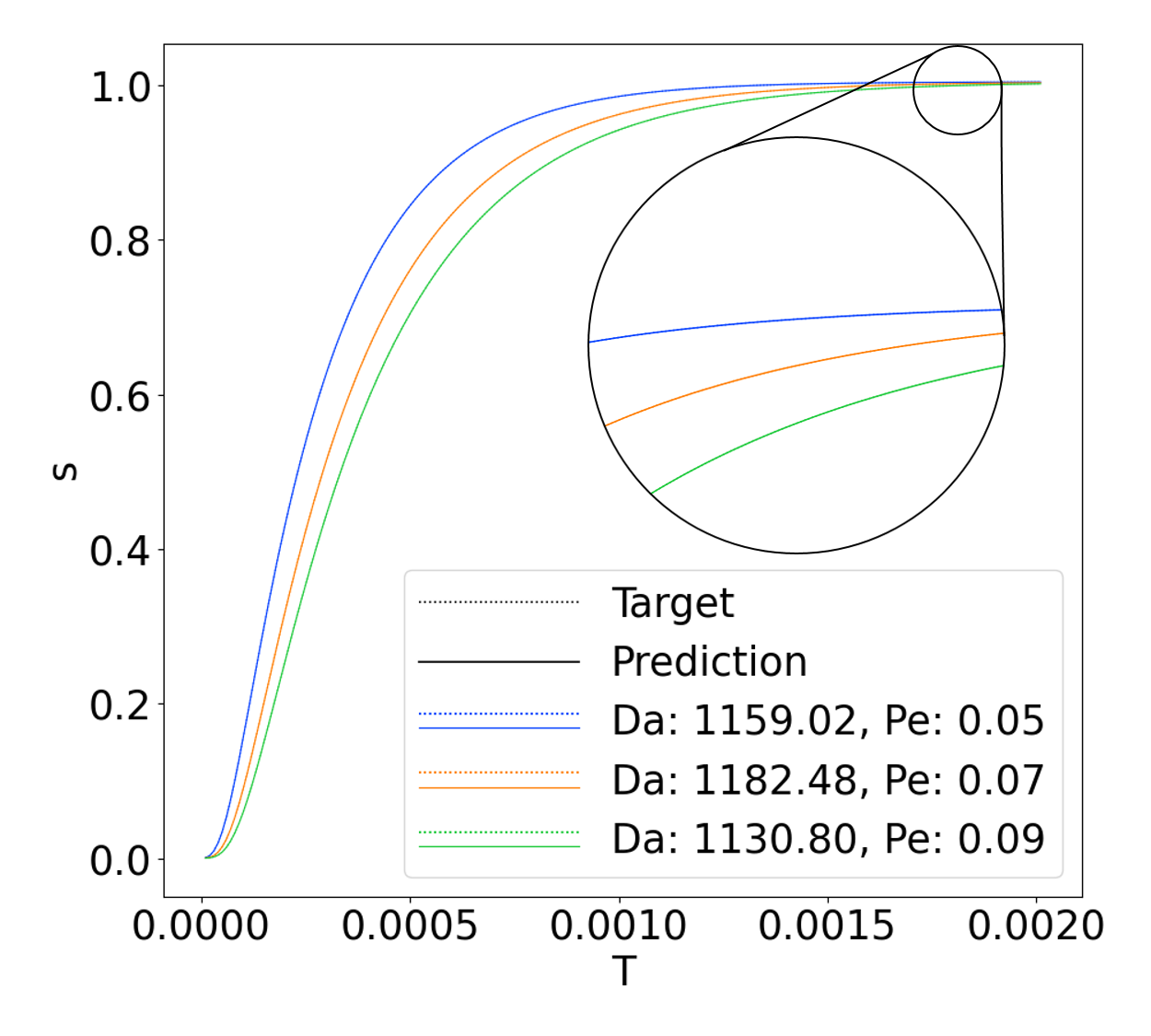}
		\caption{}
		\label{nonlin:ca}
	\end{subfigure}
	\caption{Nonlinear reaction. Breakthrough curve samples for target values and values predicted by: (a) Gaussian process, (b) neural network, (c) cross approximation.}
	\label{nonlin:bcurves}
\end{figure}

We present errors of the models in Tables~\ref{nonlin:errs_gp}, \ref{nonlin:errs_nn}, \ref{nonlin:errs_ca} for Gaussian processes, neural networks and cross approximation correspondingly. Test errors can be found in Table~\ref{nonlin:errs}. In this case we plot also samples of predicted and precomputed breakthrough curves as well (Fig.~\ref{nonlin:bcurves}). The results of the methods are similar to the results in the linear case. \replaced[id=DF]{Here}{here} it should be noted that breakthrough curves in both linear and nonlinear cases are quite similar and can sometimes coincide. This depends on the parameter values, i.e. ratio $Da/Pe$.

\begin{table}[]
	\caption{Nonlinear reaction. Results for Gaussian processes for different kernels.}
	\label{nonlin:errs_gp}
	
	\begin{tabular}{p{1.7cm}p{1.7cm}p{1.7cm}p{1.7cm}p{1.7cm}}
		\hline\noalign{\smallskip}
		\multirow{2}{*}{Kernel} & \multicolumn{2}{c}{Train error} & \multicolumn{2}{c}{Validation error}                                    \\
		\noalign{\smallskip}\cline{2-5}\noalign{\smallskip}
								& $e_{l_2}$     	   & $e_{\max}$           & $e_{l_2}$     	   			  & $e_{\max}$   \\
		\noalign{\smallskip}\svhline\noalign{\smallskip}
		RBF                     & $9.43\times 10^{-6}$ & $7.24\times 10^{-5}$ & $\mathbf{7.20\times 10^{-6}}$ & $\mathbf{2.76\times 10^{-5}}$ \\
		Matern                  & $6\times 10^{-5}$    & $0.00075$            & $2.15\times 10^{-5}$          & $0.00013$                     \\
		RQ                      & $1.12\times 10^{-5}$ & $0.00013$            & $7.91\times 10^{-6}$          & $6.42\times 10^{-5}$          \\
		\noalign{\smallskip}\hline\noalign{\smallskip}
	\end{tabular}
\end{table}

\begin{table}[]
	\caption{Nonlinear reaction. Results for neural networks for activation functions.}
	\label{nonlin:errs_nn}
	
	\begin{tabular}{p{2cm}p{1.7cm}p{1.7cm}p{1.7cm}p{1.7cm}}
		\hline\noalign{\smallskip}
		\multirow{2}{*}{\makecell{ Activation\\ function}} & \multicolumn{2}{c}{Train error}             & \multicolumn{2}{c}{Validation error}             \\
		\noalign{\smallskip}\cline{2-5}\noalign{\smallskip}
								& $e_{l_2}$     	   & $e_{\max}$           & $e_{l_2}$     	   			  & $e_{\max}$    \\
		\noalign{\smallskip}\svhline\noalign{\smallskip}
		ReLU                    & $6.91\times 10^{-5}$ & $0.00031$            & $8.48\times 10^{-5}$          & $0.00031$          \\
		Leaky ReLU              & $3.77\times 10^{-5}$ & $8.87\times 10^{-5}$ & $\mathbf{5.35\times 10^{-5}}$ & $\mathbf{0.00021}$ \\
		ELU                     & $0.00045$            & $0.00055$            & $0.00045$                     & $0.00054$          \\
		\noalign{\smallskip}\hline\noalign{\smallskip}
	\end{tabular}
\end{table}

\begin{table}[]
	\caption{Nonlinear reaction. Results for cross-approximation and different interpolation methods.}
	\label{nonlin:errs_ca}
	
	\begin{tabular}{p{2.3cm}p{1.7cm}p{1.7cm}p{1.7cm}p{1.7cm}}
		\hline\noalign{\smallskip}
		\multirow{2}{*}{Interpolation} & \multicolumn{2}{c}{Train error}            & \multicolumn{2}{c}{Validation error}                  \\
		\noalign{\smallskip}\cline{2-5}\noalign{\smallskip}
									   & $e_{l_2}$     	   	  & $e_{\max}$           & $e_{l_2}$     	   			   & $e_{\max}$  \\
		\noalign{\smallskip}\svhline\noalign{\smallskip}
		Polynomial                     & $1.89\times 10^{-6}$ & $3.64\times 10^{-6}$ & $\mathbf{1.82\times 10^{-6}}$   & $\mathbf{3.59\times 10^{-6}}$          \\
		Piecewise linear               & $4.06\times 10^{-6}$ & $9.78\times 10^{-6}$ & $4.05\times 10^{-6}$            & $9.92\times 10^{-6}$ \\
		\noalign{\smallskip}\hline\noalign{\smallskip}
	\end{tabular}
\end{table}

\begin{table}[]
	\caption{Nonlinear reaction. Test errors}
	\label{nonlin:errs}
	
	\begin{tabular}{p{3cm}p{1.7cm}p{1.7cm}}
		\hline\noalign{\smallskip}
		\multirow{2}{*}{Kernel} & \multicolumn{2}{c}{Test error}                   \\
		\noalign{\smallskip}\cline{2-3}\noalign{\smallskip}
								& $e_{l_2}$     	   & $e_{\max}$              \\
		\noalign{\smallskip}\svhline\noalign{\smallskip}
		Gaussian process        & $6.55\times 10^{-6}$   & $2.15\times 10^{-5}$            \\
		Neural network          & $4.68\times 10^{-5}$   & $0.00016$ \\
		Cross approximation     & $1.83\times 10^{-6}$   & $3.44\times 10^{-6}$  \\
		\noalign{\smallskip}\hline\noalign{\smallskip}
	\end{tabular}
\end{table}

\section{Conclusion}

We have performed a comparison study of two machine learning methods and one tensor method extended by interpolation \added{when applied to predict breakthrough curves for a catalytic filter. The training sets were obtained by simulating at pore scale 3D reactive flow using the software tool PoreChem}. All the methods show good performance, though for different regimes the order of error differs. In the \added{the} case of a strong reaction \replaced[id=DF]{the errors obtained are much smaller than in the diffusion-dominated case}{it is easier to restore the breakthrough curve\added{s}}. This may be due to the fact, that the curves changed less for \deleted{within} the chosen intervals of parameters. The methods requiring small amount of data (Gaussian processes and cross approximation) are more beneficial in this case, as numerical simulations for larger geometries require hours of running to compute just one point. In our case, it took several minutes to compute one sample.

In \added{the} case of cross approximation we have empirically shown, \added{that} the selected basis vectors \replaced{for the}{in} low-rank approximation together with \added{the} data selected by maximum volume algorithm\added{,} provide a good approximation to the considered tensor. Further interpolation of {the} basis vectors gives us a good extension to the continuous case.

Future work includes applying the considered methods for cases of more complex reactions and more complex geometries. The investigation how the intervals influences the models performances \replaced[id=DF]{is}{\replaced{should}{can} be also} a matter of further studies. Additionally, these approaches can be considered for similar tasks in other areas, e.g. decline curves in oil industry.

\bibliographystyle{plain}
\bibliography{bibliography}
\end{document}

%% file: ca_tensor.tex
    \begin{tikzpicture}

    \draw (0,0) -- (6,0);
    \draw (6,0) -- (6,6);
    \draw (6,6) -- (0,6);
    \draw (0,6) -- (0,0);
    \node (A) at (0,0) [label=left:$p_1^1$]{};
    \node (B) at (0,6.1) [label=left:$p_2^1$]{};
    \node (A1) at (0,0) [label=below:$p_1^2$]{};
    \node (C) at (6,0) [label=below:$p_2^2$]{};
    \draw[draw=mypink, fill=mypink] (2.8,0) rectangle ++(0.4,6);
    \draw[draw=mypink, fill=mypink] (0,2.8) rectangle ++(6,0.4);
    \draw[draw=mypink] (2.8,2.8) rectangle ++(0.4,0.4);
    \node (Pe) at (3,-1) {$Pe$};
    \node [rotate=90] (Da) at (-1,3) {$Da$};
    
    \draw[draw=myblue, fill=mediumturquoise] (1.6,1.6) rectangle ++(0.4,0.4);
    \draw[draw=myblue, fill=mediumturquoise] (1.6,3.6) rectangle ++(0.4,0.4);
    \draw[draw=myblue, fill=mediumturquoise] (1.6,4.5) rectangle ++(0.4,0.4);
    \draw[draw=myblue, fill=mediumturquoise] (1.6,5.2) rectangle ++(0.4,0.4);
    
    \draw[draw=myblue, fill=mediumturquoise] (3.6,1.6) rectangle ++(0.4,0.4);
    \draw[draw=myblue, fill=mediumturquoise] (3.6,3.6) rectangle ++(0.4,0.4);
    \draw[draw=myblue, fill=mediumturquoise] (3.6,4.5) rectangle ++(0.4,0.4);
    \draw[draw=myblue, fill=mediumturquoise] (3.6,5.2) rectangle ++(0.4,0.4);
    
    \draw[draw=myblue, fill=mediumturquoise] (4.8,1.6) rectangle ++(0.4,0.4);
    \draw[draw=myblue, fill=mediumturquoise] (4.8,3.6) rectangle ++(0.4,0.4);
    \draw[draw=myblue, fill=mediumturquoise] (4.8,4.5) rectangle ++(0.4,0.4);
    \draw[draw=myblue, fill=mediumturquoise] (4.8,5.2) rectangle ++(0.4,0.4);
    
    \draw [dashed, myblue]  (-0.2,1.6) -- (6.2,1.6);
    \draw [dashed, myblue]  (-0.2,2) -- (6.2,2);
    \draw [dashed, myblue]  (-0.2,3.6) -- (6.2,3.6);
    \draw [dashed, myblue]  (-0.2,4) -- (6.2,4);
    \draw [dashed, myblue]  (-0.2,4.5) -- (6.2,4.5);
    \draw [dashed, myblue]  (-0.2,4.9) -- (6.2,4.9);
    \draw [dashed, myblue]  (-0.2,5.2) -- (6.2,5.2);
    \draw [dashed, myblue]  (-0.2,5.6) -- (6.2,5.6);
    
    \draw [dashed, myblue]  (1.6,-0.2) -- (1.6, 6.2);
    \draw [dashed, myblue]  (2.0,-0.2) -- (2.0, 6.2);
    \draw [dashed, myblue]  (3.6,-0.2) -- (3.6, 6.2);
    \draw [dashed, myblue]  (4.0,-0.2) -- (4.0, 6.2);
    \draw [dashed, myblue]  (4.8,-0.2) -- (4.8, 6.2);
    \draw [dashed, myblue]  (5.2,-0.2) -- (5.2, 6.2);

    \node (i1) at (-0.4, 1.8) {$A_{i_1}$};
    \node (k1) at (-0.4, 3.0) {$A_{k_1}$};
    \node (ik) at (-0.4, 5.4) {$A_{i_{r_1}}$};
    \node (j1) at (1.8, -0.4) {$B_{j_1}$};
    \node (k2) at (3.0, -0.4) {$B_{k_2}$};
    \node (jk) at (5.0, -0.4) {$B_{j_{r_2}}$};

    \node [align=center] (init_el) at (8, 2) {Initialy\\ selected\\ elements} ;
    \node (empty1) at (6,3) {};
    \draw [-Latex] (init_el) -- (empty1.center);
    
    \node [align=center] (maxvol_el) at (8, 5.5) {Elements\\ selected\\ by MaxVol} ;
    \node (empty2) at (5.2,5.4) {};
    \node (empty3) at (5.2,3.8) {};
    \draw [-Latex] (maxvol_el.west) -- (empty2.center);
    \draw [-Latex] (maxvol_el.west) -- (empty3.center);
    \node (t) at (1.4,0.8) [label={[shift={(-0.4,-0.2)}]$t$}]{};
    \draw [dashed,-Latex] (A.center) -- (t.center);
    
    \end{tikzpicture}